# Synthesis, characterization and low temperature studies of iron chalcogenide superconductors

J. Janaki[†], T. Geetha Kumary, Awadhesh Mani, S. Kalavathi, G V R Reddy[*], G.V. Narasimha Rao[*] and A. Bharathi

Materials Science Group, Indira Gandhi Centre for Atomic Research, Kalpakkam-603102, India

[*] International Advance Research Centre for Powder Metallurgy & New Materials, Hyderabad, India

**Abstract**

We have synthesized tetragonal iron selenide and telluride superconductors through solid state reaction at 450 °C and 550 °C respectively. These synthesis temperatures have been established by optimization. Electrical resistivity and magnetic susceptibility measurements (4.2-300 K) confirm superconductivity with $T_C$ of around 8.5 K in all selenide samples of nominal composition $Fe_{1+\delta}Se$ ($\delta$=0.02-0.22). However, Scanning Electron Microscopy/Energy Dispersive Spectroscopy studies clearly indicate that the actual stoichiometric ratio of Fe : Se for all the samples synthesized is around *1:1*. Also all the samples exhibit identical phase transition temperatures from tetragonal to orthorhombic structure below ~ 100 K implying identical phase composition. The partial substitution of Se by Te leads to an enhancement of $T_C$ to ~ 13 K. The non-superconducting telluride $Fe_{1.09}Te$ exhibits a metal-insulator phase transition at ~ 82 K. Substitution studies of this telluride system by S and Si have in addition been carried out to investigate if chemical pressure induces superconductivity.

[†] Corresponding author; email: jjanaki@igcar.gov.in

## INTRODUCTION

The new iron chalcogenide layered superconducting compounds, discovered by Hsu et al. [1], have aroused fresh interest in the field of unconventional superconductors. These compounds exhibit superconductivity with $T_C \sim 8\text{-}14$ K and a considerably high critical magnetic field of 20-30T [1, 2]. Recently, $T_C$ has been raised to 27 K under application of external pressure [3]. With respect to their crystal structure, they exhibit tetragonal symmetry at room temperature with space group P4/nmm. Their structure resembles the iron arsenide based layered compounds [6] and can be described as a defect antifluorite - type layered structure. Each iron is tetrahedraly surrounded by four Se atoms forming edge shared tetrahedral layers. The Se atoms are also four coordinated and located at the apex of a pyramid with four Fe atoms on one side. Iron is in square planar coordination with neighboring Fe atoms. These compounds are reported to transform to a lower symmetry structure at low temperatures with the structural transition temperature greater than $T_C$ [4, 5]. The signature of this structural transition is also observed in electronic properties including resistivity. The electronic structure of superconducting FeSe has been investigated by the photoemission spectroscopy and is in overall agreement with the band structure calculations confirming the dominance of Fe d-state at Fermi level $E_F$ [7]. The neutron scattering studies [8] reveal a unique complex incommensurate antiferromagnetic order in the parent compound $Fe_{1+\delta}Te$, which is non-superconducting.

An open question in these systems of compounds is the correlation between superconductivity and cation/anion stoichiometry. While some reports claim that nonstoichiometry of Se in $FeSe_{1-\delta}$ of the order of 0.12-0.18 is essential for superconductivity, others observe superconductivity in the stoichometric compound itself [1, 2, 4]. Detailed structural studies by X-ray and neutron diffraction on the isostructural telluride based compounds, however, indicate that iron excess rather than tellurium deficiency is present in $Fe_{1+\delta}Te$ system wherein the excess iron is present in the octahedral interstitial positions [5]. On the other hand, for the $Fe_{1+\delta}Se$ system such iron defects have not been clearly identified. The recent phase diagram study is consistent with the system being essentially stoichiometric with a very narrow homogeneity range and iron excess, even if present, cannot exceed 0.03 [9]. In order to investigate possible existence of composition-structure-$T_C$ correlations, we have studied selenides with the nominal composition $Fe_{1+\delta}Se$ ($\delta$=0.02 to 0.22). The pseudo-binary system $Fe_{1+\delta}Se_{1-x}Te_x$(x=0.5) has also been synthesized in the present study and found to have higher $T_C$ than the pure selenide. The telluride system $Fe_{1+\delta}Te$ does not show superconductivity but exhibits a metal insulator transition. With the idea of searching for new superconducting compositions in the telluride system, partial substitutions of Si and S for Te have been attempted. The results of our studies on the above systems are reported.

**EXPERIMENTAL DETAILS**

Polycrystalline samples of selenides and tellurides of nominal composition $Fe_{1+\delta}Se$ ($\delta$= 0.02-0.22), $Fe_{1.18}Se_{0.5}Te_{0.5}$, $Fe_{1.09}Te$, $Fe_{1.09}Te_{1-x}A_x$ (A= Si, S and x=0.05 and 0.2) were prepared by solid state reaction technique. The constituent elements, namely, Fe (99.99% pure), Se (99.9% pure) and Te (99.9% pure) were taken in the stochiometric ratio. The Fe powder was hydrogen reduced at 750°C prior to use. The pellets made out of the well mixed powders were sealed in evacuated quartz tubes and allowed to react at temperatures of 450 °C and 550 °C for the selenide and telluride respectively for three days. Our preliminary attempts towards the synthesis of tetragonal FeSe/Te at temperatures between 680-700 °C as reported in literature[1,2] gave rise to multiphase samples with $Fe_3O_4$, formed due to oxidation of Fe under the high temperature reaction conditions. The oxide content was minimized by employing a lower synthesis temperature and using Fe purified by hydrogen reduction. The crystal structure was characterized by X-ray diffraction (XRD) measurements using a STOE diffractometer. Electrical resistivity and magnetic susceptibility measurements were carried out in a dipstick-cryostat in the range 4.2-300 K using the four probe technique and the *a.c.* mutual inductance technique respectively. Micro-structural analysis was carried out by using XL30, SEM/EDS microscope.

**RESULTS AND DISCUSSION**

Fig.1 shows the XRD patterns corresponding to our initial attempts at the synthesis of tetragonal FeSe at 680°C as reported in literature [1, 2]. As can be seen, these samples $Fe_{1+\delta}Se$ ($\delta$=0 and 0.09), hence forth designated as HT1 and HT2 respectively, contain predominantly hexagonal phase ($P6_3/mmc$) instead of the desired tetragonal FeSe (P4/nmm). In addition, minority phase of $Fe_3O_4$, is formed due to oxidation of Fe under the high temperature reaction conditions. The XRD patterns of $Fe_{1+\delta}Se$ ($\delta$=0.02, 0.14 & 0.22) prepared at 450°C hence forth designated as LT1, LT2 and LT3, are shown in Fig. 2. The results of XRD phase analyses indicate that under the present conditions of synthesis all the three compositions have the same degree of phase purity and are all around 95% single phase with the tetragonal (P4/nmm) structure. The remaining 5% impurities present in the samples include mainly Fe, $Fe_3O_4$ and hexagonal ($P6_3/mmc$) FeSe. All the XRD patterns in Fig. 2 have been indexed to the P4/nmm space group with a Figure of Merit F(10) > 60 and lattice parameters obtained are listed in Table 1. The lattice parameter values are in rough agreement with literature [1, 2] and do not shift with composition for $Fe_{1+\delta}Se$ ($\delta$=0.02 - 0.22) .

Compositional analysis was carried out using Energy Dispersive Spectroscopy in different regions of the pellets and found to be uniform. Table 2 presents the results of a typical analysis for three pellets of nominal compositions $Fe_{1.02}Se$, $Fe_{1.14}Se$ and $Fe_{1.22}Se$ (LT1-3). As can be seen from the table, the Fe:Se ratio for all the pellets is around *1:1*. It is worth noting that upon material balance considerations one would expect that if all the samples contain stoichiometric *1:1* FeSe, a considerable amount of Fe based impurity of ~2%, 14% and 22% should be present for the samples $Fe_{1.02}Se$, $Fe_{1.14}Se$ and $Fe_{1.22}Se$ respectively. However, this is not detected in our XRD analysis. This may be present in the form of $Fe_3O_4$/Fe at the grain boundaries with coherence lengths less than the detectable limit of XRD. Also oxygen impurity >10% has been identified in EDS analysis of the samples $Fe_{1+\delta}Se$ ($\delta$=0.14 - 0.22), whose concentration increases with excess iron content. However, a detailed grain boundary analysis has not yet been undertaken.

Figures 3a – c show the temperature dependent resistivity, $\rho(T)$, of $Fe_{1+\delta}Se$ (for $\delta$=0.02-0.22) samples prepared at $450^0$ C. The positive slope of $\rho(T)$ indicates that these samples exhibit metallic behavior. The normal state resistivity shows *T*-linear behavior between $T_C$ and 100 K, beyond which it deviates from linearity (Figs.3a - c). A *T*-linear behavior at low temperature, rather than the $T^2$ dependence characteristic of a Fermi liquid, is well known to arise for High $T_C$ cuprates and is regarded as a signature of a non-canonical Fermi liquid system [11]. The deviation from the linearity at ~ 100 K is reported to occur in this system [12] due to structural transition from tetragonal to orthorhombic symmetry. All the samples exhibit onset $T_C$ ~ 8.5 K. For comparison, the $\rho(T)$ of $Fe_{1+\delta}Se$ ($\delta$=0 and 0.09) samples prepared at $680^0$ C (HT1 and HT2) are shown in fig. 3d. It can be seen that $\rho(T)$ of these samples is different from the LT1-3. Fig. 4 exhibits region near the superconducting transition for all the samples. It should be mentioned that in contrast to the LT1-3 which show comparatively sharper transition with zero resistivity, the poor quality HT1-2 samples prepared at higher temperature exhibit broad transition with non-vanishing resistivity in the case of HT1.

The magnetic susceptibility $\chi(T)$ studies carried out on the selenide samples of nominal composition $Fe_{1+\delta}Se$ ($\delta$=0.02-0.22) are presented in Fig. 5. It is observed that an increase in the value of $\delta$ leads to a suppression of superconductivity (i.e. decrease in superconducting volume fraction).

In the inset of this figure $\chi(T)$ of low quality samples (HT1-2) is also shown for comparison. These samples show a clear signature of the well known Verwey transition at 125K [10] which can be taken as a fingerprint of the $Fe_3O_4$ impurity. This impurity signal has been considerably reduced upon optimizing the synthesis temperature to 450°C for the selenides. Interestingly there is significant increase in the diamagnetic signal (corresponding to superconductivity) in samples LT1-3.

In what follows, we present the results of our investigation on the superconducting properties of the pseudo binary compound $Fe_{1+\delta}Se_{0.5}Te_{0.5}$ and the binary compound $Fe_{1+\delta}Te$. Both of these compounds are isostructural to the selenides. Fig. 6 a and Fig. 7 a depicts the XRD of the above compounds. The lattice constants '$a$' & '$c$' deduced from XRD are found to increase from $Fe_{1+\delta}Se \rightarrow Fe_{1+\delta}Se_{0.5}Te_{0.5} \rightarrow Fe_{1+\delta}Te$ as shown in Table 1. This is in accordance with the higher ionic radius of Te compared to Se. The telluride is evidently less pure compared to the selenide (see Fig. 7 a) because of the higher synthesis temperature. The contribution of $Fe_3O_4$ is found to be > 10% from XRD. Therefore, resistivity anomaly seen at 125 K (see Fig. 5 inset) is attributed to the Verwey transition on account of the presence of $Fe_3O_4$, which is observed in all the telluride samples. The formation of $Fe_3O_4$ leads to a depletion of iron concentration in the telluride samples and hence promotes the formation of $FeTe_2$ impurity phase as seen in XRD patterns (Fig. 7). It is suggested that stringent and rigorous synthesis techniques to prevent inclusion of oxygen, like, adding an oxygen getter or carrying out synthesis in high purity argon atmosphere rather than vacuum or double sealing of the evacuated quartz tubes will help in synthesizing a phase pure material. The $\rho(T)$ behavior of the pseudo-binary compound $Fe_{1.18}Se_{0.5}Te_{0.5}$ is shown in Fig 6 b. The $T_C$ can be raised from 8.5 K to 13 K in going from the compound $Fe_{1+\delta}Se \rightarrow Fe_{1+\delta}Se_{0.5}Te_{0.5}$ (Fig. 6 b and 6 b inset ).

With the idea of searching for new superconducting compositions in the telluride system we have attempted Si and S substitution at Te site in $Fe_{1+\delta}Te$. The substitution of Si and S, which have smaller ionic radii than Te, would be equivalent to applying chemical pressure. This would be equivalent to applying external pressure, which is known to enhance $T_C$ in FeSe. The compounds synthesized include $Fe_{1+\delta}Te_{1-x}A_x$ (A= Si ,S and x=0.05 and 0.2). Fig. 7 shows the representative XRD patterns for A= Si, S and x=0.2. The results of crystal structure and phase analysis of the above series by XRD indicate that no changes in lattice parameters upon Si doping but only a decrease in the intensity of $Fe_3O_4$ compared to the pristine compound. The latter has roughly the same $Fe_3O_4$ content as the S doped compound. Hence, it is clear that Si does not enter the lattice under the synthesis conditions but only getters the oxygen from $Fe_3O_4$. On the other hand, S doping leads to a significant contraction in the lattice parameters (Fig. 7 and Table1) confirming the formation of substituted compounds. Fig. 8 shows the $\rho(T)$ behavior of $Fe_{1+\delta}Te_{1-x}A_x$ (A= S, with x=0.05 and 0.2 respectively) as well as pristine $Fe_{1+\delta}Te$. It is seen that $Fe_{1+\delta}Te$ is not superconducting but exhibits a metal-insulator transition near 82 K. This is in accordance with literature [5, 14] wherein it is known that the metal-insulator transition (MIT) is accompanied with a structural transition from tetragonal → monoclinic form, which in turn associated with a transformation from paramagnetic to anti-ferromagnetic behavior. All the S doped samples exhibit superconductivity with onset $T_C \sim$ 8.5 K (see Fig. 8). In contrast to pristine telluride, the S doped samples exhibit insulating behavior in the normal state over the entire temperature range and exhibit higher resistivity (Fig 8) indicating that the samples are perhaps granular. It is interesting to note that the structural anomaly corresponding to the MIT at 82 K in the pristine telluride is pushed to lower temperature around 30 and 24 K with S substitution in $Fe_{1.09}S_{0.05}Te_{0.95}$ and $Fe_{1.09}S_{0.2}Te_{0.8}$ respectively and becomes more feeble leading to the occurrence of insulator - insulator transitions. This is apart from the expected anomaly observed in the resistivity curves at ~ 125 K for all the telluride compounds due to the presence of $Fe_3O_4$ as discussed earlier. It should be noted that MIT in the pristine telluride appears concomitantly with the structural and magnetic transition, which is also known to shift to lower temperature from 82 - 62 K [15] with increasing iron excess. The electron doping effect arising due to the excess iron perhaps shifts the transition to lower temperature. However, a larger dramatic shift of this structural/magnetic transition is observed upon S substitution due to internal chemical pressure and this correlates with the occurrence of superconductivity. Hence, it may be concluded that the absence of superconductivity in the pristine telluride can be correlated with the presence of the higher energy change, first order structural / magnetic phase transition occurring at higher temperature in pristine telluride than the S doped tellurides, which suppresses $T_C$ plausibly due to long range magnetic ordering. Magnetization and specific heat studies have been planned on this system to study the origin of this phase transition and its relationship to superconductivity in further detail. Towards the completion of our work a report on S substitution has appeared in literature and the results are in agreement with our study [16].

## SUMMARY

Our present studies of the XRD, SEM/EDS and low temperature (4.2-300 K) electrical resistivity and magnetic susceptibility of the iron selenide superconducting system is consistent with the fact that the tetragonal superconducting phase of iron selenide is essentially stoichiometric FeSe and has neither appreciable selenium deficiency nor iron excess. The excess iron is added to account for the inadvertent oxidation of Fe to $Fe_3O_4$, which perhaps precipitates at the grain boundaries. The $T_C$ of the selenide is enhanced from 8.5 K to 13 K by substitution of Te for Se. However, the pure telluride is non-superconducting and exhibits a metal-insulator phase transition at~ 82 K. The suppression of this MIT by doping S at Te site gives rise to superconductivity with $T_C$ ~ 8.5K.

**Figure Captions**

Fig. 1. The XRD patterns of $Fe_{1+\delta}Se$ ($\delta$ = 0 and 0.09) synthesized at 680 °C; each data is normalized to maximum peak height and the data shifted vertically for clarity.

Fig. 2. The XRD patterns of $Fe_{1+\delta}Se$ ($\delta$ = 0.02, 0.14 and 0.22) synthesized at 450 °C; each data is normalized to maximum peak height and the data shifted vertically for clarity.

Fig. 3. (a, - c) Variation of electrical resistivity in temperature range of 4.2 - 150K of $Fe_{1+\delta}Se$ ($\delta$=0.02-0.22) and (d) $Fe_{1+\delta}Se$ ($\delta$=0 and 0.09) synthesized at 450 °C and 680 °C respectively. The straight line shows the linear variation in the temperature range from $T_C$ to ~ 100 K. Arrows indicate the anomaly at the phase transition.

Fig. 4. Variation of electrical resistivity of $Fe_{1+\delta}Se$ ($\delta$=0.02-0.22) and $Fe_{1+\delta}Se$ ($\delta$=0 and 0.09) synthesized at 450 °C and 680 °C respectively, in the region near the superconducting transition (4.2 - 18 K ).

Fig. 5. AC magnetic susceptibility, '$\chi_{ac}$' versus temperature of $Fe_{1+\delta}Se$ ($\delta$=0.02-0.22) showing the diamagnetic superconducting transition. Inset shows the behavior of the 680 °C synthesized, $Fe_{1+\delta}Se$ ($\delta$=0 and 0.09). The asterisk indicates the magnetic signal corresponding to $Fe_3O_4$ impurity.

Fig. 6. (a) The XRD pattern of $Fe_{1.18}Se_{0.5}Te_{0.5}$. Inset shows the electrical resistivity versus temperature plot near the superconducting transition for $Fe_{1.18}Se_{0.5}Te_{0.5}$.

Fig. 7. XRD patterns of $Fe_{1.09}Te_{0.8}A_{0.2}$ (A=Te, Si, S); each data is normalized to maximum peak height and the data shifted vertically for clarity.

Fig. 8. Variation of electrical resistivity of $Fe_{1.09}Te_{1-x}S_x$ (x = 0, 0.05 and 0.20) between 4.2 and 300K. The resistivity data near the superconducting transition is shown in the inset. The arrows indicate the resistivity anomaly at the phase transition. The asterisks indicate the signal corresponding to $Fe_3O_4$ impurity.

Fig1:

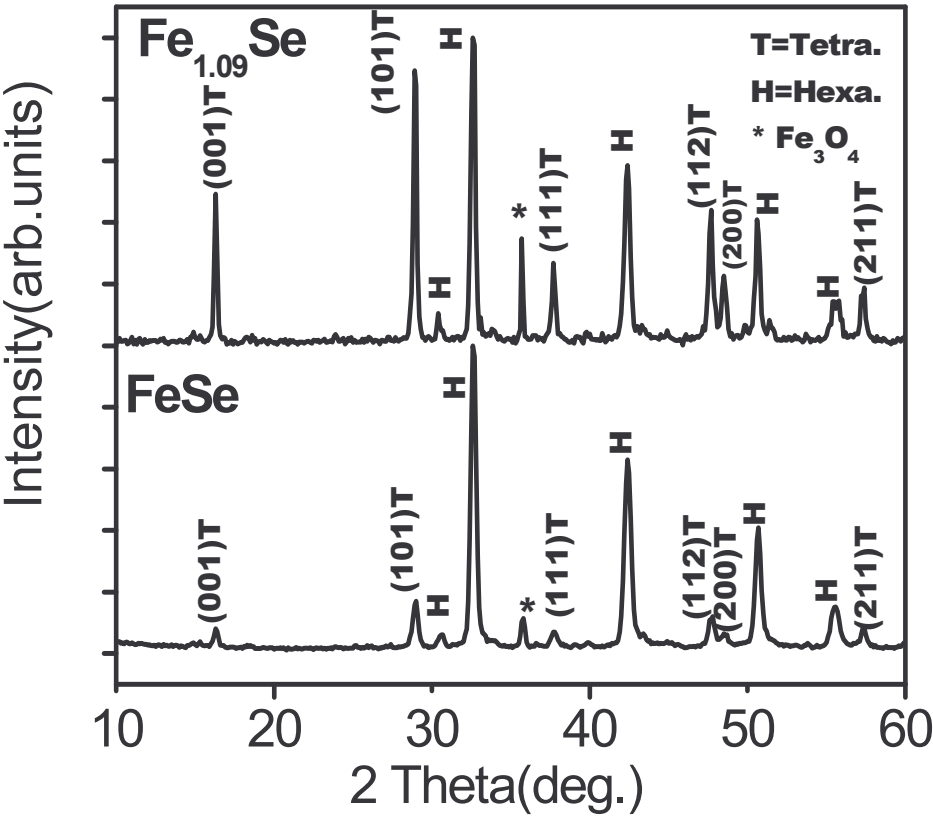

Fig2:

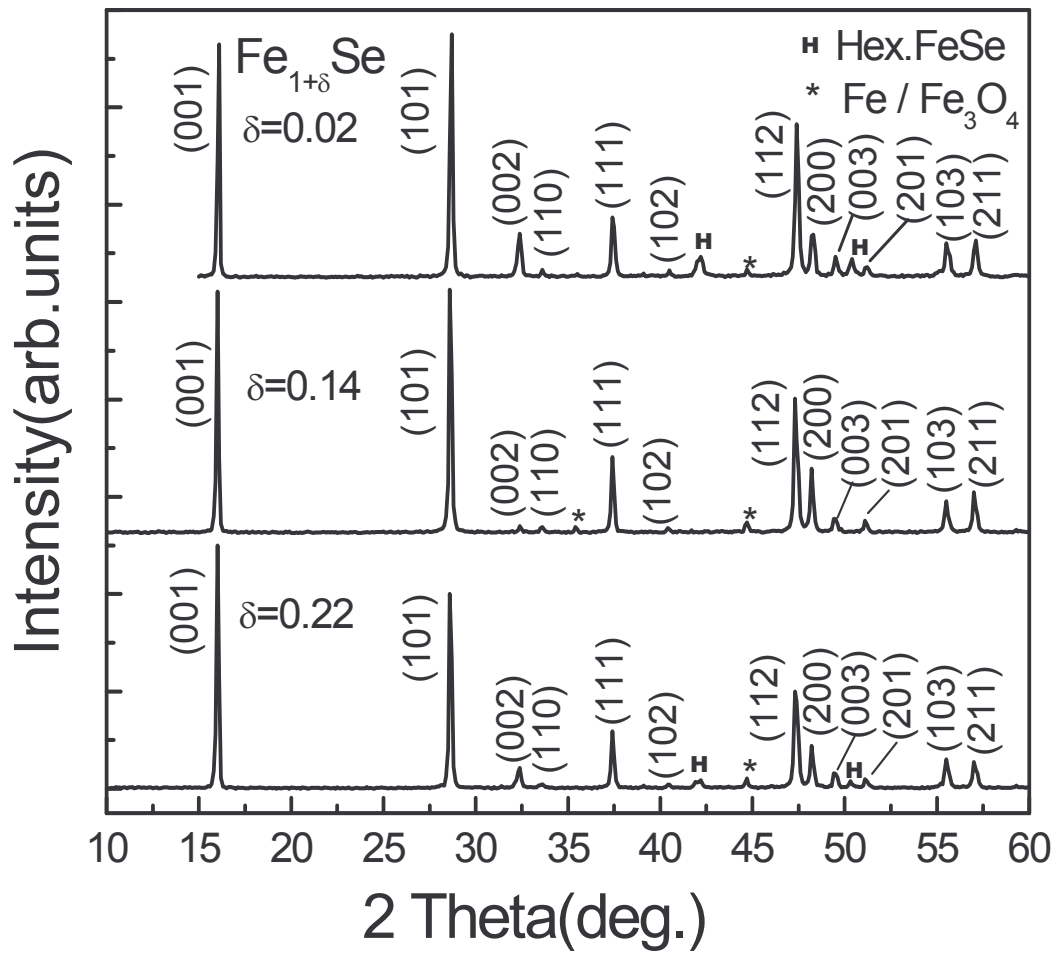

Fig3:

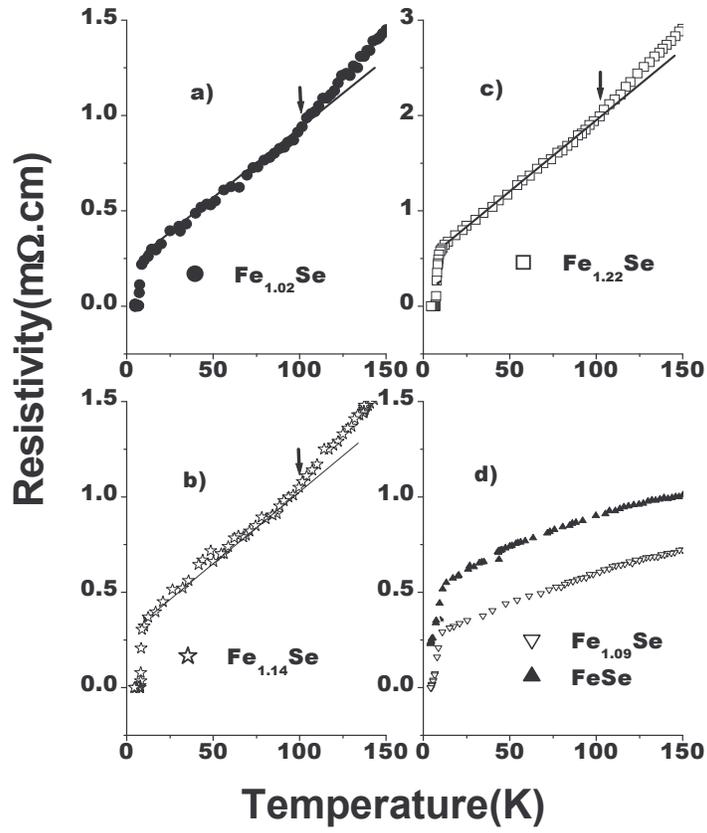

Fig4:

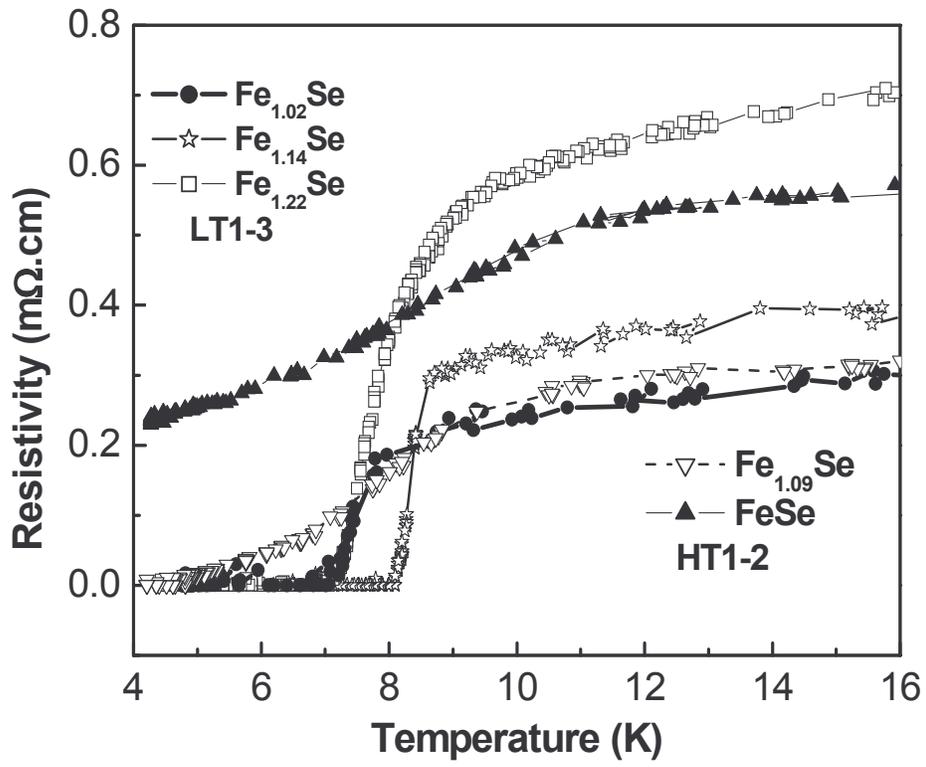

Fig5:

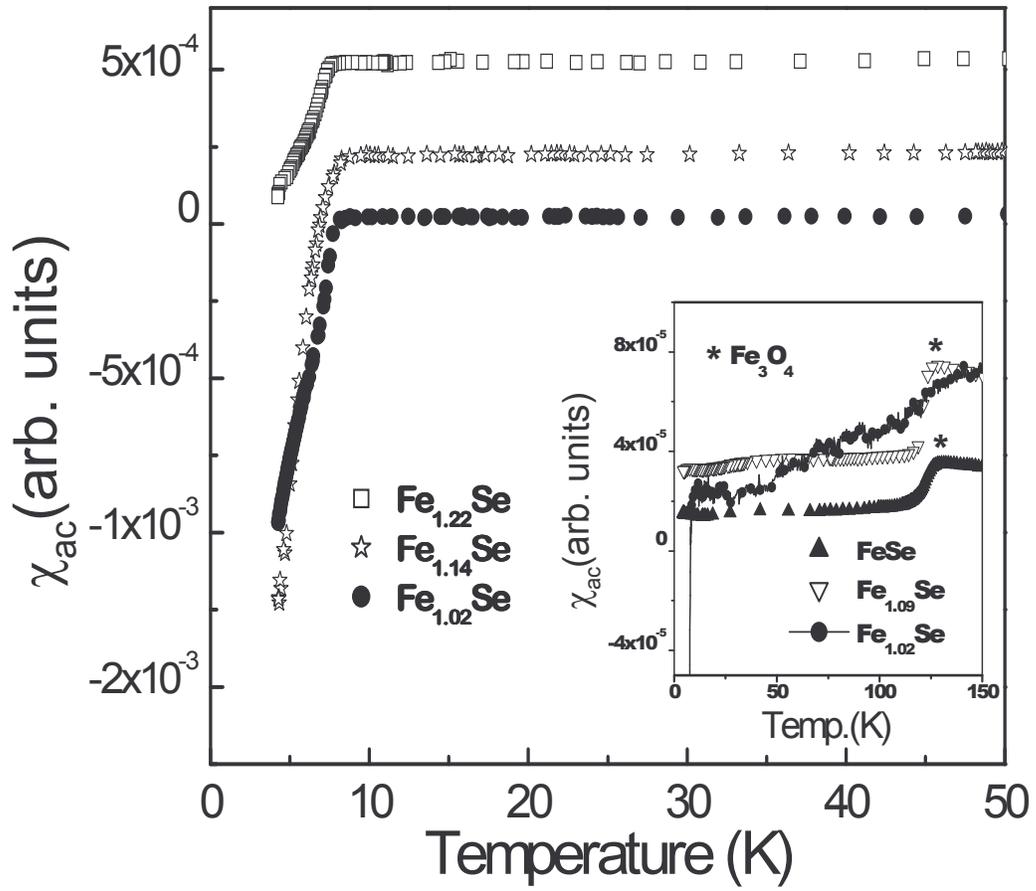

Fig6:

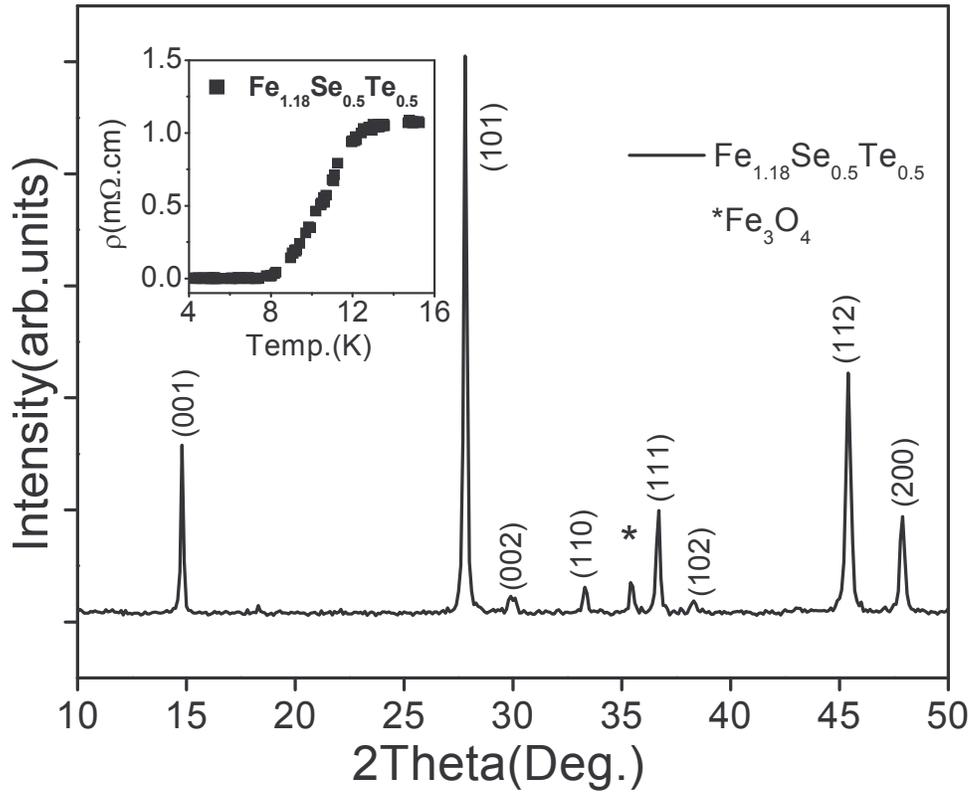

Fig7:

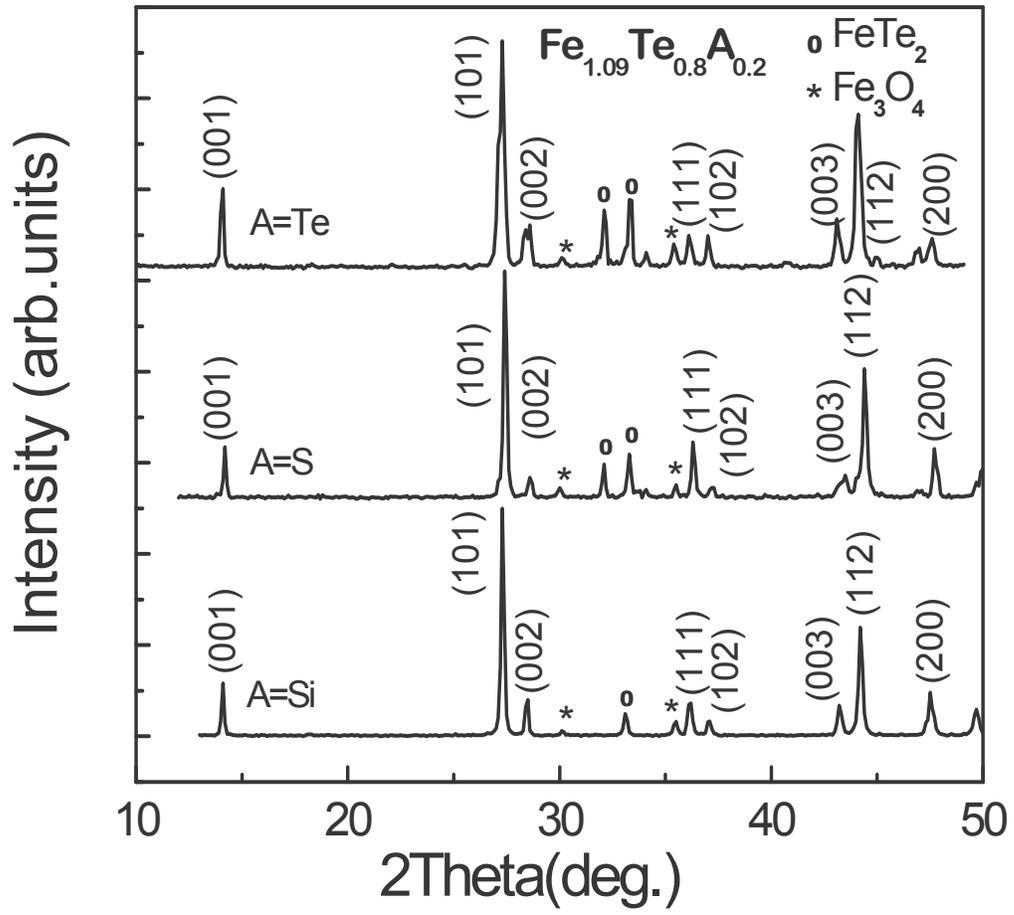

Fig8:

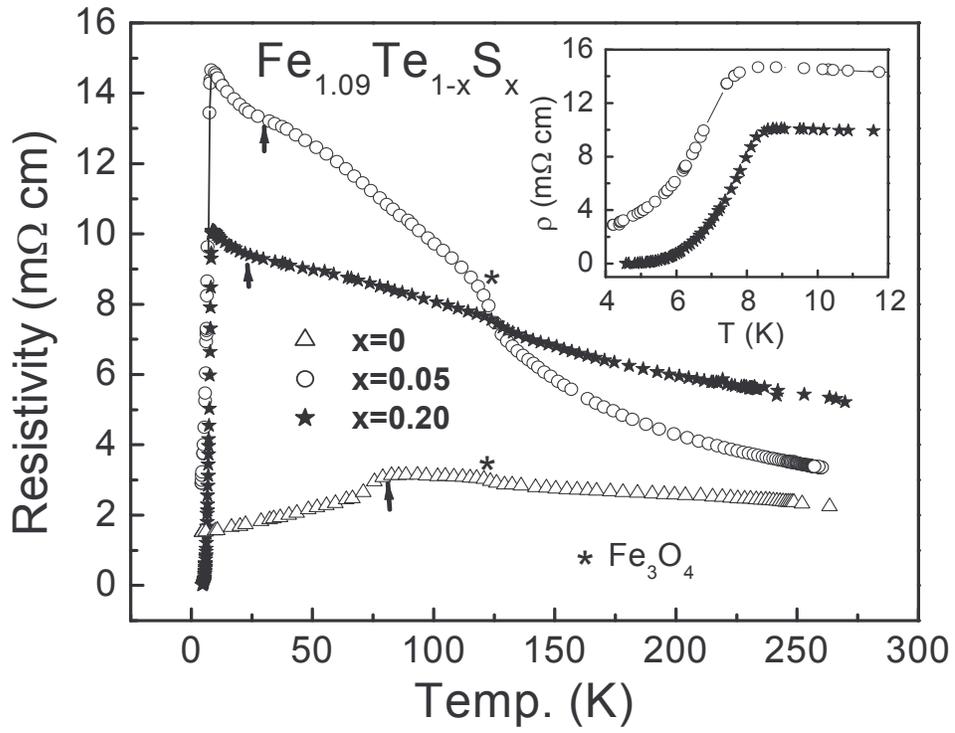

Table 1: Lattice parameters estimated from XRD analysis for the samples studied *

| Iron Selenide System | Tetragonal lattice parameters(A°) | | Iron Telluride System | Tetragonal lattice parameters(A°) | |
|---|---|---|---|---|---|
| | "a" | "c" | | "a" | "c" |
| $Fe_{1.02}Se$ | 3.771 | 5.523 | $Fe_{1.09}Te$ | 3.823 | 6.280 |
| $Fe_{1.14}Se$ | 3.772 | 5.521 | $Fe_{1.09}Si_{0.2}Te_{0.8}$ | 3.822 | 6.272 |
| $Fe_{1.22}Se$ | 3.773 | 5.523 | $Fe_{1.09}S_{0.2}Te_{0.8}$ | 3.807 | 6.240 |
| $Fe_{1.18}Se_{0.5}Te_{0.5}$ | 3.797 | 5.971 | | | |

* Note that the compositions referred to are nominal compositions

Table 2: Compositions of iron selenide pellets in three different regions (1) (2) & (3) determined from SEM/EDS analysis for nominal compositions as indicated

| Sample | Atom% Fe (1) | Atom% Se (1) | Atom% Fe (2) | Atom% Se (2) | Atom% Fe (3) | Atom% Se (3) |
|---|---|---|---|---|---|---|
| $Fe_{1.02}Se$ | 48.1 | 51.9 | 48.1 | 51.9 | 48.0 | 52.0 |
| $Fe_{1.14}Se$ | 49.1 | 50.9 | 48.1 | 51.9 | 48.5 | 51.5 |
| $Fe_{1.22}Se$ | 48.8 | 51.2 | 49.1 | 50.9 | 48.6 | 51.4 |